\documentstyle[12pt]{article}

\begin{document}
\centerline{\bf Theoretical Approach to Alignment Phenomenon}
\ \\[.5cm]

\centerline{\bf I. Royzen}
\ \\[.3cm]

\centerline{P.N. Lebedev Physical Institute of Russian Academy of Sciences}
\centerline{53 Leninsky Prospect, 117924 Moscow, Russia}
\ \\[.2cm]

\begin{abstract}
An explanation of the puzzling alignment effect observed in
cosmic ray experiments is suggested
\end{abstract}

   Few years ago the observation has been made [1] in cosmic ray experiments
that the alignment of the main energy fluxes along a straight line in target
(transverse) plane exceeds significantly the background level. More precisely,
at superhigh energies of initial particle ($E_{0} \geq 10^{4} \mbox{ TeV}$) the
secondary particle superfamilies detected by deep lead X-ray emulsion chamber
appeared to be situated almost along straight line in target plane (Fig.1). The
coplanar scattering of such a type was so surprising that an attempt has been
made to revise the result but instead they were confirmed with much better
confidence level [2]. The analysis of the alignment effect for 74 high energy
$\gamma$-families induced by hadrons above and within the
chamber has been carried out. Their energies energies are selected to be
$\sum E_{\gamma}=100\div5000\ \mbox{ TeV}$ (hadron energies being restored,
accounting that the energy of induced   $\gamma$-family is about $1/3$ of the
hadron energy it is originated from). This analysis suggested that superfamily
production happened predominantly rather low above the chamber (at the altitude
$H\simeq 2 \mbox{km}$, since it seemed that nuclear-electromagnetic cascade
development would blur alignment, if several interactions contributed). It
confirmed a coplanar scattering and scaling-like fragmentation spectrum of
energy distinguished cores. The alignment parameter $\lambda$,
$$
-0.5 \leq \lambda =
\frac{\sum_{i\not=j\not=k}^{m} \cos(2\phi_{ijk})}{m(m-1)(m-2)}\leq 1 ,
$$
\newpage
is used as the alignment criterion where $m$ stands for a number of centers
of highest energy and $\phi_{ijk}$ is the angle between the two-dimensional
vectors
$\vec k_i$ and $\vec k_j$ in target plane, an event being recognized to have
alignment, if $\lambda\geq 0.6$ . Actually, events with $m=4$ were chosen only
because of too high statistical background for $m=3$  and rather poor
statistics
for $m\geq 5$. The threshold-like behavior of the effect has been observed: no
alignment at $\gamma$-family energies $\sum E_{\gamma}\le 100 \mbox{ TeV}$ ,
then its gradual increase within energy range
$ 100\mbox{TeV}\le\sum E_{\gamma}\le 500 \mbox{TeV}$ to manifest itself finally
in (20-40)\% of total number of
events. 14 events with $\sum E_{\gamma}\geq 500 \mbox{ TeV}$ have been
observed,
exhibiting most striking alignment structure ($\lambda\geq 0.8$ ).
Core transverse momentum $p_T$  was estimated by rough relation
$p_T H \simeq E_0R $, $R$  being the distance of a spot from the interaction
axis. The mean ratio of value of maximal relative core transverse momentum to
its normal to the alignment line projection (in target plane) $k_T$
is $<p_T>/<k_T>\simeq 10$. No other peculiarities of alignment events compared
to "usual" cascade have been noticed.

The first attempt of theoretical consideration of the above alignment
phenomenon
has been made by F. Halzen and D.A. Morris [3], whose approach was based on the
assumption that semihard gluon jets is a feature of all events at energies
above $10^{4} \mbox{ TeV}$. It was shown that within this approach the cosmic
ray observations were associated probably with the jet alignment in three-jet
events observed already in the collider experiment [4].

I would like to suggest an alternative treatment which makes it possible to
understand many features of cosmic ray
alignment observations quite naturally, including the threshold-like
energy behavior and fraction in extensive atmospheric showers as well as the
typical projections of core transverse momenta to the alignment line and
normal to it, and allowing for events with more, than four cores aligned,
that have been extracted recently from cosmic ray data [5]. The main point of
the approach under consideration is that the alignment events are assumed
to be associated with semihard double inelastic diffraction
(SHDID) of hadrons [6]. Let us trace them step-by-step.

\section{ Total cross section of SHDID.}

   In the accordance with the conventional Regge-Gribov approach, the
one-Pomeron
contribution to the differential cross section of SHDID  can be expressed at
$s \gg M^2_{1,2} \gg Q^2_{T} \gg 1 \mbox{ GeV}^2$ in the form:

\begin{eqnarray}\label{1}
\frac{d\sigma^{0}_{DD}}{d Q_T^2}&=&
\frac{\sigma_t r^2(Q_T^2)}{16\pi}
\int_{Q_T^2/\epsilon}^{\epsilon s}\frac{dM^2_1}{M^2_1}
\int_{Q_T^2/\epsilon}^{sQ_T^2/M_1^2}\frac{dM^2_2}{M^2_2}
\left(2\frac{sQ_T^2}{M_1^2M_2^2}\right)^{2(\alpha_P -1)}\\
 &=& \frac{4^{\alpha_P -1}\sigma_t r^2(Q_T^2)}{32\pi(1- \alpha_P)}
\ln\frac{\epsilon^2 s}{Q_T^2}\nonumber
\end{eqnarray}

where $\sqrt{s}, \sigma_t, Q_T, M_1$ and $M_2$ are CMS total
interaction energy,
total cross section, transverse momentum transferred, and invariant masses of
final diffractively excited
states respectively, $r(Q_T^2)$ is three-Pomeron vertex
and $\alpha_P\equiv \alpha_P(Q_T^2)$ is the Pomeron
trajectory; the parameter $\epsilon=max (M^2_{1,2}/s) \simeq 0.05$ is to be
chosen to single out diffraction processes from other ones [7].
Since  the mean slope of the Pomeron trajectory is the only dimensional
parameter which can be responsible for the decrease of the function $r(Q_T^2)$
as $Q^2$ is increased, the domain where $r(Q_T^2)$ is expected to be nearly
constant is estimated as
$Q_T^2\leq(\overline{\alpha'_P})^{-1}$ where $\overline{\alpha'_P}$ is an
effective mean value of the derivative ${\alpha}_P'$ there which is reasonably
evaluated to be $\overline{\alpha'_P}\leq (0.1 - 0.2)\mbox{ GeV}^{-2}$. It is
why this domain is expected to be remarkably large, from $Q_T^2= 0$  to
$Q_T^2 \simeq 10 \mbox{ GeV}^2$ or even larger
(it has been observed long ago by comparison of the elastic and single
inelastic diffraction differential cross sections
that $r(Q^2_T)\simeq \mbox{const}$ at $Q^2_T\leq 1.5\mbox{ GeV}^2$ [7],
wherefrom, in particular,
a rather slow $Q_T$-dependence of double inelastic diffraction differential
cross section at $Q^2_T \leq 1,5 \mbox{ GeV}^2$ follows). The double
inelastic diffraction is the only type of hadron
interaction which is expected to exhibit such slow transverse momentum
dependence. At still larger values of squared 4-momentum transferred Pomeron is
expected to be dissolved to its constituents [6] that
begin to interact independently, so that the "normal" QCD regime
$\alpha^2_S/Q^{-4}$ is to be approached gradually. In what follows the
logarithmic dependence on $Q_T$ and rather ambiguous but definitely slow
decrease of
$\alpha_P(Q_T^2)$ in the right-hand side of eq.(1) are accounted on the average
as $Q^2_T\to \overline{Q^2_T} \mbox{ and } \alpha_P(k^2_T)\to
\overline{\alpha_P(k^2_T)}$ . The rough estimate of screening corrections
to the one-Pomeron SHDID scattering amplitude $A_{DD}^0$ associated with
diagrams depicted in Fig.3 shows that
\begin{equation}\label{2}
A_{DD}\simeq \frac{A_{DD}^0}{1+2\frac{\sigma_{el}}{\sigma_{t}}}
\end{equation},
$A_{DD}$ being the corrected amplitude.
It is reasonable to adopt ${\sigma_{el}}/{\sigma_{t}}\simeq 0.2$ and enhance
the above correction (i.e.,to multiply the denominator in eq.(2)) by the
phenomenologically
approved (for forward elastic scattering amplitude) factor about 1.5,
accounting
the shadowing by the inelastic intermediate states. Then the corrected SHDID
amplitude is expected to be $A_{DD} \simeq 0.55A^0_{DD}$
and the corresponding differential cross section is
$$
\frac{d\sigma_{DD}}{dk_T^2}\simeq
0.3\frac{d\sigma_{DD}^0}{dk_T^2}\ .
$$
After integration of eq.(1) over the
region $Q_T^2\leq (\overline{\alpha'_P})^{-1}$ one obtains the total cross
section of SHDID

\begin{equation}\label{3}
\sigma_{DD}\simeq \frac{0.3\ 4^{\overline{\alpha_P}}\sigma_t r^{2}(0)}
{128\pi \overline{\alpha'_P}(1-\overline{\alpha_P})}
\ln (\overline{\alpha'_P}{\epsilon^2s})
\end{equation}

If one chooses a reasonable values $\overline{\alpha_P}\simeq 0.5$,
$\overline{\alpha'_P}\simeq 0.15 \mbox{ GeV}^{-2}$
and the experimental value of $r_0$, $r_0\simeq 0.8\mbox{ GeV}^{-1}$, then
the fraction of SHDID is expected to be ${\sigma_{DD}}/{\sigma_{t}}
\simeq 0.04; 0.07;$ and 0.10 at $s = 10^{5}; 10^{6}$
and $10^{7} \mbox{ GeV}^{2}$ respectively.
It can be several times less or larger,
since the above estimate is rather rough, but its smooth logarithmic
threshold-like energy increase is independent of the choice of parameters.

\section{ Transverse (target) plane structure of events.}

It seems reasonable to expect that hadronization of diffractively excited
final states produced by SHDID is dominated by mechanism of string rupture as
shown in Fig.4, string been formed between scattered colored hadron constituent
(quark, diquark or gluon) and remnant of the same hadron. Any alternative
string
configuration would be unfavorable since it implies formation of some strings
of a very high energy (it is worthy to mention that diffractively produced
state associated with target particle was always out of the game in cosmic ray
experiments under discussion because it is never seen within the area of
observation; it is why the projectile inelastic diffraction only is thought of
throughout the paper). At the same time, transferred momentum
$Q_{T}\simeq 3\mbox{ GeV}$ is insufficiently large for the fragmentation
mechanism of hadronization to prevail.
Let us consider the above string in its own CMS and adopt that secondary
particle rapidity
and transverse momentum distributions in Pomeron-proton interaction is similar
to that in real hadron one at CMS energy $M_{1} (or M_{2})$ (as to the
rapidity distribution, it is supported by the well known result of UA4
Collaboration [8]). Since what is
observed is nothing else, than transverse plane projection of the picture which
is resulted from its rupture, it becomes obvious that the typical ratio of
a secondary transverse momentum projection normal to
reaction plane (i.e., to the plane of draft) to "transverse momentum string
length" (i.e. to LS relative transverse momentum of leading particles
oppositely directed in string CMS ) is
about $\frac{k_{T}\sqrt2}{Q_{T}}$ where $<k_{T}>\simeq
300\mbox{MeV}$ is mean transverse momentum of secondaries in hadron
interactions
, and mean leading particle energy is experimentally proved to be about
half of incident particle one. At $Q_{T}\simeq 3\mbox{GeV}$ this ratio is about
0.13.

\section{ Comparison to the experimental data.  }

The only point what remains to be discussed to compare the above consideration
to the experimental data is an obvious estimate of the role of
atmospheric cascade. Since the atmosphere thickness above the
altitude where the calorimeter is mounted corresponds to about 3.5 nuclear
mean free paths, the probability of at least one SHDID collision is about $1
-(1 - \frac{\sigma_{DD}}{\sigma_{T}})^{3.5} \simeq 0.3$ at $s = 10^{7} \mbox{
GeV}^{2}$. If it does happen, then the subsequent soft collisions can not,
most probably, blur essentially the target plane picture it initiates,
especially for energy distinguished cores. It is why the additional
assumption suggested by experimenters [2] seems to be not necessary, that
alignment is caused by some peculiarities of the lowest nuclear collision above
the chamber only. At the same time, the threshold-like dependence of alignment
on core energies is associated, may be, with the violating role of nuclear
cascade.
   Thus, the main puzzling experimental features of alignment phenomenon,
namely,
the fraction of alignment events about (20-40)\% and the ratio of mean value of
normal to reaction plane projection of core transverse momentum to
maximal value of core relative transverse momenta ($\simeq0.1$) (string
"half-thickness" to its "length" in transverse momentum space) are compatible
qualitatively with the above theoretical consideration (30\% and 0.13
respectively), if one adopts that each core is originated (due to
electromagnetic cascade) from a hadron created along with string rupture. The
threshold-like dependence of SHDID cross section on interaction energy can
elucidate why the phenomenon
has not been noticed at lower energies (especially,
accounting a poor statistics and other ambiguities of cosmic ray experiments).
However, this point as well as some other features of the phenomenon, such as
its threshold-like dependence on core energies, core energy distribution, their
energy sequence along the alignment line, etc., needs both the enrichment of
statistics and MC simulation of cascade and SHDID collisions themselves
(especially, accounting that hadrons of different masses can be produced at the
end of string and along its length) which are in progress. Unfortunately,
it is rather questionable, whether an attempt to observe the alignment
phenomenon will be undertaken in accelerator experiments soon.

   I would like to thank A. Capella, E. Feinberg, J. Tran Thanh Van, and
especially A. Managadze for many fruitful discussions.

   The work is supported, in part, by Russian Foundation for
Fundamental Researches and International Science Foundation.  

\newpage
 \centerline{Figure captions}

\vspace{2.0cm}

\begin{description}
\item{Fig.1.} The example of target plane picture with energy
distinguished cores for event with alignment, $\lambda = 0.95$; figures stand
       for energy in TeV (already multiplied by factor 3 for
hadrons);\hspace{1cm}and\hspace{1cm}or \hspace{1cm}stand for electromagnetic
halo and hadrons of high energy respectively.  Other particles of the family
are marked as\hspace{1cm}($\gamma$ -quanta) and \hspace{1cm}(hadrons).
\item{Fig.2.} One-Pomeron exchange approximation to SHDID. Wavy lines refer to
       Pomeron exchange, $M_1$ and $M_2$ are invariant masses of diffractively
excited states, Q is 4-momentum transferred, $r$ is triple-Pomeron vertex
       function.
\item{Fig.3.} Typical diagrams, accounting screening corrections.
Notation is the same as in Fig.2.
\item{Fig.4.} The scheme of final state hadronization by string rupture
mechanism.
\end{description}
\end{document}